\documentclass[preprint,superscriptaddress,nofootinbib,preprintnumbers,amsmath,amssymb]{revtex4-1}
\usepackage{latexsym}
\usepackage{graphicx}
\usepackage{verbatim}
\usepackage{amsthm}
\usepackage{float}
\usepackage{amsmath}
\usepackage{booktabs}
\usepackage{graphicx,subfigure}
\usepackage{epstopdf}
\usepackage{indentfirst}
\usepackage{epsfig}
\usepackage{epsfig,subfigure,psfrag}
\usepackage{dcolumn}
\usepackage[T1]{fontenc}
\usepackage{lmodern}
\usepackage{bm}
\usepackage{slashed}
\usepackage{cancel}
\usepackage{color}
\usepackage{tabularx}
\usepackage{hyperref}
\hypersetup{
    colorlinks=true,       
    linkcolor=blue,          
    citecolor=blue,        
    filecolor=blue,      
    urlcolor=blue           
}

\makeatletter
\newcommand{\Spvek}[2][r]{%
  \gdef\@VORNE{1}
  \left(\hskip-\arraycolsep%
    \begin{array}{#1}\vekSp@lten{#2}\end{array}%
  \hskip-\arraycolsep\right)}

\def\vekSp@lten#1{\xvekSp@lten#1;vekL@stLine;}
\def\vekL@stLine{vekL@stLine}
\def\xvekSp@lten#1;{\def\temp{#1}%
  \ifx\temp\vekL@stLine
  \else
    \ifnum\@VORNE=1\gdef\@VORNE{0}
    \else\@arraycr\fi%
    #1%
    \expandafter\xvekSp@lten
  \fi}
\makeatother

\newcommand{\be}{\begin{equation}}
\newcommand{\ee}{\end{equation}}
\newcommand{\beq}{\begin{eqnarray}}
\newcommand{\eeq}{\end{eqnarray}}

\newcommand{\bea}{\begin{eqnarray}}
\newcommand{\eea}{\end{eqnarray}}

\newcommand{\GeV}{{~\rm GeV}}
\newcommand{\gev}{{~\rm GeV}}

\newcommand{\hplus}{\ensuremath{{H^+}}}

\newcommand{\hhp}{\ensuremath{{H^0}}}
\newcommand{\ahp}{\ensuremath{{A^0}}}
\newcommand{\hsm}{\ensuremath{{H_{_\text{SM}}}}}

\newcommand{\tth}{\ensuremath{t\bar{t}\hsm}}

\newcommand{\lsim}{\begin{array}{c}\,\sim\vspace{-26pt}\\< \end{array}}
\newcommand{\gsim}{\begin{array}{c}\sim\vspace{-26pt}\\> \end{array}}

\newcommand{\tanb}{\ensuremath{\rm tan \beta}}
\newcommand{\fb}{{\rm fb}}

\begin{document}

\title{Charged Higgs Signals in $t\,\overline{t}\,H$ Searches}

\author{Daniele S. M. Alves}
 \email{spier@nyu.edu}
 \affiliation{Center for Cosmology and Particle Physics, Department of Physics, New York University, New York, NY 10003}
 \affiliation{Department of Physics, Princeton University, Princeton, NJ 08544}

\author{Sonia El Hedri}
 \email{elhed001@uni-mainz.de}
 \affiliation{PRISMA Cluster of Excellence \& Mainz Institute for Theoretical Physics, Johannes Gutenberg University, 55099 Mainz, Germany}

\author{Anna Maria Taki}
 \email{amt543@nyu.edu}
 \affiliation{Center for Cosmology and Particle Physics, Department of Physics, New York University, New York, NY 10003}

\author{Neal Weiner}
 \email{neal.weiner@nyu.edu}
 \affiliation{Center for Cosmology and Particle Physics, Department of Physics, New York University, New York, NY 10003}

\date{\today}

\begin{abstract}

New scalars from an extended Higgs sector could have weak scale masses and still have escaped detection. In a Type I Two Higgs Doublet Model, for instance, even the charged Higgs can be lighter than the top quark. Because electroweak production of these scalars is modest, the greatest opportunity for their detection might come from rare top decays. For mass hierarchies of the type $m_t>m_\hplus>m_{\ahp,\,\hhp}$, the natural signal can arise from top quark pair production, followed by the decay chain $t \rightarrow b\hplus$, $\hplus \rightarrow  W^{+(*)} \phi^0$, $\phi^0\rightarrow b\overline{b},\,\tau^+\tau^-$, where $\phi^0=\ahp,\,\hhp$. These final states largely overlap with those of the Standard Model $t\overline{t}\hsm$ process, and therefore can potentially contaminate $t\overline{t}\hsm$ searches. We demonstrate that existing $t\overline{t}\hsm$ analyses can already probe light extended Higgs sectors, and we derive new constraints from their results. Furthermore, we note that existing excesses in $t\overline{t}\hsm$ searches can be naturally explained by the contamination of rare top decays to new light Higgses. We discuss how to distinguish this signal from the Standard Model process.

\end{abstract}

\maketitle

\tableofcontents
\newpage

\numberwithin{equation}{section}
\renewcommand\theequation{\arabic{section}.\arabic{equation}}
\renewcommand*{\thefootnote}{\fnsymbol{footnote}}
\section{\label{sec:intro} Introduction}
So far, LHC searches have not provided conclusive signs of new particles, nor significant deviations from Standard Model predictions. Generic limits on new colored particles are particularly severe, with squarks and gluinos from Supersymmetry already tightly constrained if lighter than 1.3~TeV and 1.9~TeV, respectively \cite{ATLAS:2016kts}.

Exclusion limits on direct production of new electroweak particles, in contrast, have not been as dramatic. Mass limits on charginos produced via electroweak interactions, for instance, do not extend beyond 400-500 GeV \cite{CMS:2016gvu}. The reason for this is straightforward: at a $p$-$p$ machine such as the LHC, electroweak cross sections are simply much smaller than strong cross sections. Because of that, the best means to search for new electroweak states is often in cascade decays of copiously produced colored particles. This approach however relies on the existence of heavier colored particles within the reach of the LHC. As this possibility becomes more and more remote, the only realization of this scenario that can be concretely studied are rare top quark decays, which unfortunately cannot probe new particles heavier than about $170$ GeV.

The range below the top quark mass is nonetheless a well motivated region to search for states beyond the Standard Model (BSM). One possibility of great interest would be rare top decays to charged Higgs bosons, $t \rightarrow b\, \hplus$. With the $t\overline{t}$ cross section at the 13 TeV LHC being over 800 pb, even small branching ratios $\text{Br}(t \rightarrow b\, \hplus)\sim\mathcal{O}(10^{-3})$ would yield a $H^\pm$ production rate at the $\mathcal{O}$(pb) level or higher. Since direct electroweak production of these states ranges from $\mathcal{O}(30-200)~\fb$ at 13 TeV, the rate from top quark decays could easily dominate.

This scenario has not gone without scrutiny. Indeed, there is a variety of independent constraints on new charged scalars lighter than the top quark. In {\it Type II}\;  Two Higgs Doublet Models (2HDM) in particular, flavor changing observables, most notably $b\rightarrow s\gamma$, exclude $m_{H^\pm} \lsim 580 \gev$ \cite{Enomoto:2015wbn,Misiak:2017bgg}, absent cancellations. Flavor bounds are highly model dependent, however. To contrast, in {\it Type I}\;  2HDMs, constraints from $b\rightarrow s\gamma$ are mild at best, requiring only that $\tan\beta \gsim (1.5-2)$ for $m_\hplus<m_t$. Thus, at least for Type I 2HDM scenarios, there is a compelling case to consider direct searches for new Higgs states at colliders, in particular in rare top decays.


A critical point to consider here is that, in a Type I 2HDM, one of the Higgs doublets is fermiophobic. This can drastically alter the phenomenology of the charged Higgs relative to a Type II scenario. In particular, if a lighter neutral scalar $\phi^0 \,(= A^0,\,H^0)$ exists, the charged Higgs will dominantly decay as $\hplus \rightarrow \phi^0\, W^{+(*)}$, and the familiar fermionic decays (e.g., $\hplus \rightarrow \tau^+ \nu,\,c\,\bar{s},\, t^*\,\overline{b}$) will be suppressed.
In this region of parameter space, our process of interest will be:
\beq
\label{ttsignal}
p\,p~\rightarrow~ t\,\bar t~&\rightarrow&~ (bW^+)(\bar{b}\,H^-)\nonumber\\
&\rightarrow&~ (bW^+)(\bar{b}\;W^{-(*)}\phi^0)
\eeq
with $\phi^0$ dominantly decaying as
\be
\label{phidecay}
\phi^0\rightarrow b\,\bar{b}\,,\;\tau^+\tau^-,
\ee
if $m_{\phi^0}\lsim110~\text{GeV}$. The resulting final state, with rates of $\mathcal{O}$(pb) or higher, can lead to a large signal contamination in searches for the SM Higgs boson produced in association with top quark pairs, $t\bar{t}\hsm$, whose SM cross section is about $0.5~\text{pb}$. Indeed, existing and future measurements of the $t\bar{t}\hsm$ cross section can be used to constrain the Type I 2HDM signatures in (\ref{ttsignal}), (\ref{phidecay}). A more exciting prospect would be to explain recent excesses in existing $t\bar{t}\hsm$ searches as due to contamination from rare top decays to charged Higgses. While the significance of current excesses is mild, upcoming results with more data will lead to a clearer picture of the excess pattern, would it persist. 

The layout of this paper is as follows: in Sec.~\ref{sec:model} we briefly review Type I 2HDMs and describe the charged Higgs phenomenology in the light mass region. Direct and indirect bounds on the relevant region of parameter space are reviewed in Sec.~\ref{sec:constraints}. In Secs.~\ref{sec:colliders}, \ref{sec:discussion},  we discuss how $t\bar{t}\hsm$ searches can be used to constrain charged Higgs production, and describe the degree to which the claimed excess in various channels can be explained by this model. Finally, in Sec.\ref{sec:conclusion} we discuss the implications of these results and comment on future searches that might help better constrain this light region of Type I 2HDMs.

\newpage

\section{Model and Signals}
\label{sec:model}
In a type I 2HDM, one of the Higgs doublets, $H_2$, is fermiophobic, and all fermion masses stem from Yukawa couplings to $H_1$:
\be
\mathcal{L}_{\text{yukawa}}=H_1\,Q\,Y_u U^c + H_1^\dagger\,Q\,Y_d D^c + H_1^\dagger\,L\,Y_\ell E^c~+~\text{h.\,c.}
\ee
The scalar potential can be generically parameterized as \cite{Gunion:1989we}:
\begin{eqnarray}
V_{\text{scalar}}&=&\lambda_1\left(|H_1|^2-v_1^2\right)^2\,+~\lambda_2\left(|H_2|^2-v_2^2\right)^2\nonumber\\
&+&\lambda_3\left(  (|H_1|^2-v_1^2) +  (|H_2|^2-v_2^2)  \right)^2 \label{eq:higgspot}\\
&+&\lambda_4\left(  |H_1|^2|H_2|^2  - |H_1^\dagger H_2|^2   \right) \nonumber\\ 
&+&\lambda_5\left(   \text{Re}(H_1^\dagger H_2) - v_1v_2    \right)^2 \,+~ \lambda_6\left(\text{Im}(H_1^\dagger H_2)\right)^2\,,\nonumber
\end{eqnarray}
where both doublets, $H_1$ and $H_2$, have hypercharge $Y=1/2$, and for simplicity we assume that CP is conserved and all parameters in (\ref{eq:higgspot}) are real.

Conventionally, the mass eigenstates of this theory are parameterized by two angles; $\alpha$, the mixing angle between the CP-even neutral states, and $\beta$, defined as $\text{tan}\beta \equiv v_1/v_2$:
\beq
\Spvek{H_1^0; H_2^0} ~=~ \Spvek{v_1; v_2}~+~\frac{1}{\sqrt{2}} R_\alpha \Spvek{H^0_\text{light}~ ;H^0_\text{heavy}}~+~\frac{i}{\sqrt{2}}R_{\beta}\Spvek{G^0;A^0}\,,
\eeq
\beq
\Spvek{H_1^\pm; H_2^\pm} ~=~ R_\beta \Spvek{G^\pm;H^\pm}\,,
\eeq
where
\begin{equation}
	R_\alpha = \left(\begin{matrix}
	~\cos \alpha & \sin \alpha \\
	-\sin \alpha & ~\cos \alpha 
\end{matrix} \right)~~,~~~~
R_\beta = \left(\begin{matrix}
	-\sin \beta & ~\cos \beta \\
	~\cos \beta & \sin \beta 
\end{matrix} \right).
\end{equation}

Usually, the ``SM-like'' Higgs (corresponding to the state discovered at the LHC with mass $m_h=125$ GeV) is the lighter CP-even neutral scalar, $H^0_\text{light}$. 
That does not need to be the case though, and in principle the SM-like Higgs could be the heavier CP-even scalar, $H^0_\text{heavy}$. Since we are interested in both regimes, we will adopt the following, more generic parameterization \cite{Alves:2012ez}
\begin{equation}
\Spvek{H_1^0; H_2^0} ~=~ \Spvek{v_1; v_2}~+~\frac{1}{\sqrt{2}} R_{\beta+\delta} \Spvek{H_{_\text{SM}}^0;\hhp~}~+~\frac{i}{\sqrt{2}}R_{\beta}\Spvek{G^0;\ahp}\,,
\end{equation}
where
\begin{equation}
	R_{\beta-\delta} ~=~ \left(\begin{matrix}
	\sin (\beta -\delta)& -\cos (\beta - \delta) \\
	\cos (\beta -\delta)& ~\sin (\beta -\delta)
\end{matrix} \right).
\end{equation}
Here, $\delta$ is a parameter that describes the deviation from the alignment limit. If the SM-like Higgs corresponds to the lightest CP-even scalar, $\delta$ is defined by $\delta \equiv \beta - \alpha - \pi/2$. Conversely, if the SM-like Higgs corresponds to the heaviest CP-even scalar, then $\delta \equiv \beta - \alpha$. The advantage of this parameterization is that $\delta$ quantifies the deviation from a SM-like Higgs, and there is no discontinuity in our description of fields as the mass hierarchy changes. That is, $H_{_\text{SM}}^0$ is always the SM-like state, and $H^0$ is always the state with suppressed couplings to fermions.



In terms of the mass eigenstates, the Yukawa couplings can be re-written as:
\beq
\label{scalarYukawaL}
\mathcal{L}_{\text{yukawa}}~=~\sum_{f}~~ &&\xi_{_{H_\text{SM}}} \frac{m_f}{v}\, H^0_{_\text{SM}} ff^c  \;+~  \xi_{_H} \frac{m_f}{v}\, H^0 ff^c\quad \nonumber \\
+~  && i\,\xi_{_A}^f\; \frac{m_f}{v}\, A^0 ff^c \;+~ \xi_{_A}^f\; \frac{m_f}{v}\, \sqrt{2}\, U_{ff^{\prime}}\, H^\pm f f^{\prime\,c}\;~+~~\text{h.\,c.}\,,~~
\eeq
where $v=246$~GeV, $U_{ff^{\prime}}$ is a CKM matrix element if $f,f^{\prime\,c}$ are quarks, and $U_{ff^{\prime}}=1$ if $f,f^{\prime\,c}$ are leptons. Moreover,
\beq
\label{scalarYukawaScaling}
\xi_{_{H_\text{SM}}}=\cos\delta-\frac{\sin\delta}{\tanb}~,~~~\xi_{_H}=\sin\delta+\frac{\cos\delta}{\tanb}~,~~~\xi_{_A}^u=-\xi_{_A}^{d,e}=\frac{1}{\tanb}.
\eeq

Likewise, the EW symmetry breaking couplings of scalars to vector bosons are given by:
\beq
\mathcal{L}_{\phi VV}~=~\zeta_{\phi} \,\frac{2\,m_W^2}{v}\,\phi^0\, W^+W^- +~   \zeta_{\phi}\, \frac{m_Z^2}{v}\,\phi^0\, ZZ \;,
\label{eq:scalarcouplings1}
\eeq
where $\phi^0$ generically denotes $H^0_{_\text{SM}}$, $H^0$, $A^0$, and,
\beq
\zeta_{_{H_\text{SM}}}=\cos\delta~,~~~\zeta_{H}=\sin\delta~,~~~\zeta_{A}=0.
\label{eq:scalarcouplings2}\eeq

\begin{figure}[t]
\includegraphics[width=0.55\textwidth]{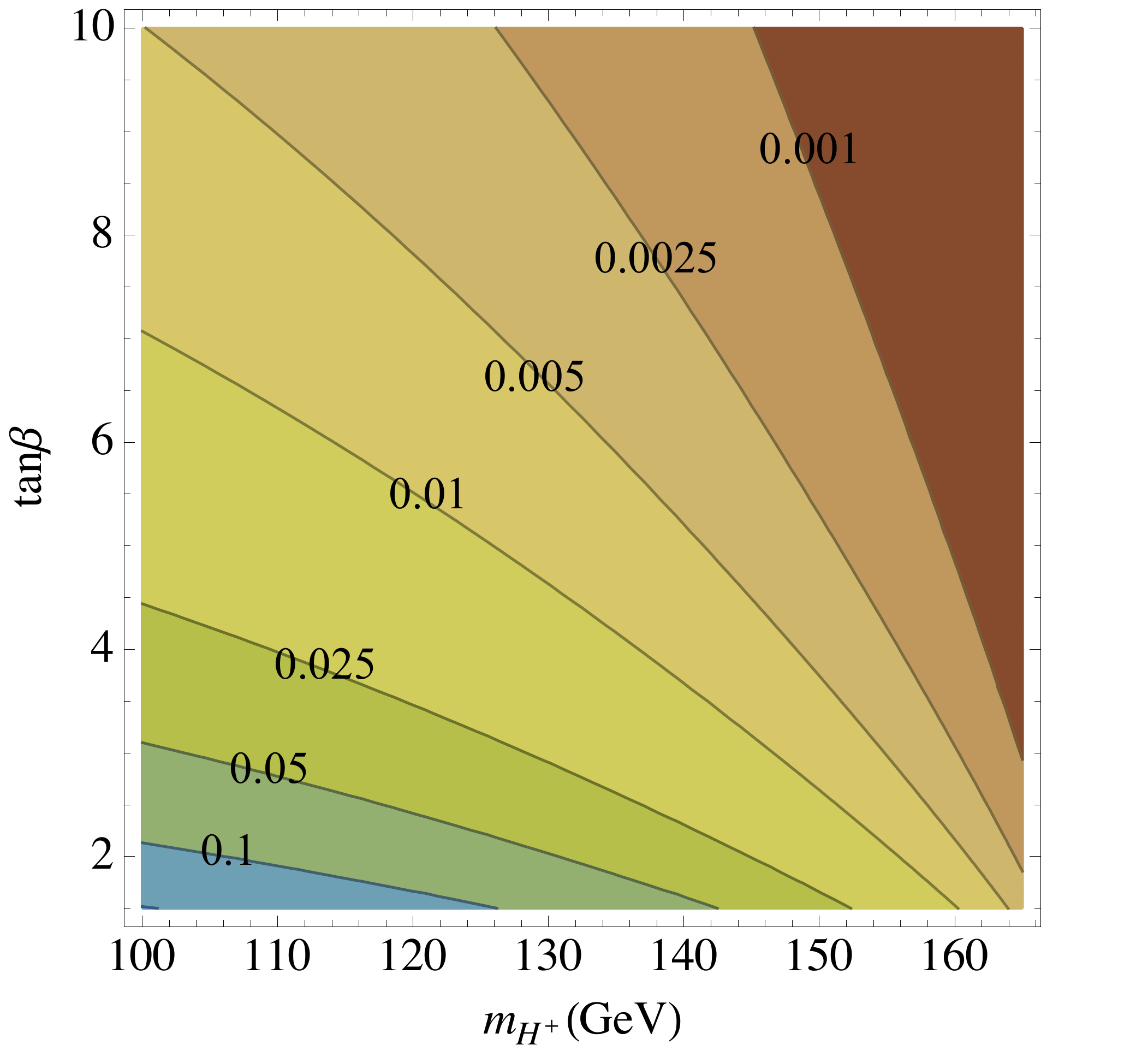}
	\caption{\label{fig:topBR} Contours of $\text{Br}(t\rightarrow b\,\hplus)$ in a Type I 2HDM, as a function of $m_{H^\pm}$ and $\tanb$.}
\end{figure}

In Type I 2HDMs, unlike Type II, the couplings of $A^0$ and $H^\pm$ to fermions are suppressed by $\tanb$, as shown in (\ref{scalarYukawaL}) and (\ref{scalarYukawaScaling}), and so are the $H^0$ yukawa couplings in the alignment limit $\delta\rightarrow 0$.

As we are especially interested in $H^\pm$ production from top decays,
we show in Fig.~\ref{fig:topBR} the branching ratio $\text{Br}(t\rightarrow b\,\hplus)$ as a function of $m_{H^\pm}$ and $\tanb$. Note that even at large $\tanb \sim \mathcal{O}(10)$, branching ratios of $\mathcal{O}(10^{-3})$ are possible, while at low $\tanb \lsim 3$, the top branching ratio to $H^\pm$ can reach the few percent level, $\text{Br}(t\rightarrow b\,\hplus)\sim\mathcal{O}(3\%) \;-$  roughly the limit where tension might arise with measurements of the top pair cross section $\sigma_{t \bar t}$ \cite{Czakon:2011xx,Peters:2015kka,Andrea:2013qoa,Schilling:2012dx,ATLAS:2012dpa,ATLAS:2012fja,CMS:2012dya,CMS:2014gta,LHCtopWG}.

\subsection{Charged Higgs Decays}

The decay patterns of the charged Higgs can vary dramatically across the parameter space of 2HDMs, significantly impacting the experimental strategies to search for this state.

\begin{figure}[t]
\includegraphics[width=0.7\textwidth]{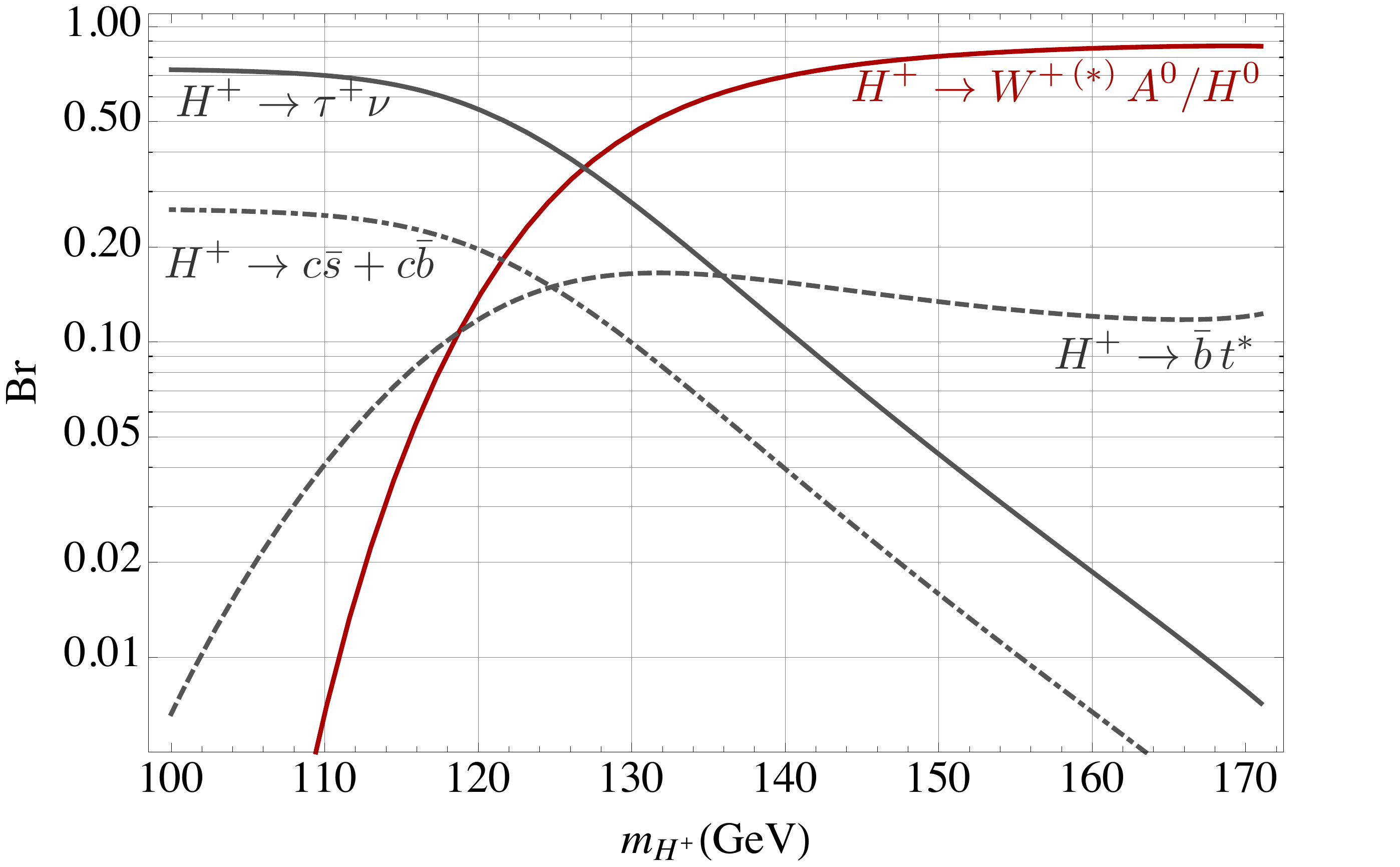}
	\caption{\label{fig:hplusdecays}Charged Higgs branching ratios in a Type I 2HDM, assuming $m_{A^0}=m_{H^0}=100$ GeV, $\tanb=3$, and $\delta=0$.}
\end{figure}


Due to the large popularity of Type II 2HDMs, the most thoroughly explored $H^\pm$ decays, both in phenomenological studies as well as in experimental efforts, have been the fermionic modes. This is due to the $\tanb$ enhancement of the $H^\pm$ couplings to down type quarks and leptons, causing the $\tau\nu$, $cs$ and $b\,t^*$ modes to dominate the $H^\pm$ branching ratios. ATLAS \cite{Aad:2014kga,Aaboud:2016dig} and CMS \cite{CMS:2014cdp,CMS-PAS-HIG-16-031} have extensively searched for signatures of $t\rightarrow b\,(H^+\rightarrow \tau^+\nu)$, and placed upper bounds on the overall top branching ratio to this final state at the sub-percent level. Analogous searches \cite{Khachatryan:2015uua,CMS-PAS-HIG-16-030} for the $H^+\rightarrow c\bar s,\,c\bar b$ channels have also set percent-level bounds on the corresponding branching ratios.

The same constraints are applicable in Type I 2HDMs if the only kinematically open channels for $H^\pm$ decays are light fermions, which is the case if $m_{H^\pm}\lsim m_\hhp,m_\ahp$. However, if the spectrum contains a lighter neutral scalar with a large $H_2$ component, such as $A^0$ or $H^0$, the mode $\hplus\rightarrow W^{\pm(*)}\,A^0/H^0$ would naturally dominate the $H^\pm$ branching ratio, even with the 3-body phase space suppression of an off-shell $W^{*}$. This is due to the unsuppressed gauge couplings that mediate this decay mode, and the large suppression of the competing fermionic modes, stemming from the smallness of the Yukawa couplings and from the $1/\tanb$ dependence of the $H^\pm$ coupling to fermions.
This effect is illustrated in Fig.~\ref{fig:hplusdecays}, where we have set $m_{A^0}=m_{H^0}=100$ GeV, $\tanb=3$, and $\delta=0$. The suppression of fermionic modes is even more pronounced for larger values of $\tanb$, or lower masses of $A^0$ or $H^0$. This qualitatively different phenomenology of charged Higgs decays has been previously noted in phenomenological studies 
\cite{Djouadi:1995gv,Akeroyd:1998dt,Kling:2015uba,Coleppa:2014cca,Coleppa:2014hxa,Arhrib:2016wpw,Bechtle:2016kui}, but to the best of our knowledge, no dedicated experimental analysis has explicitly searched for these signatures. A critical question is then: could such a particle have contaminated studies of other SM or BSM processes? And if so, what constraints could existing searches place on these particular charged Higgs signals?

\section{Constraints on a Light Higgs Sector}
\label{sec:constraints}
While new scalars with sizeable couplings to SM fermions or gauge bosons are subject to constraints from LEP, Tevatron, and LHC data, additional Higgs bosons with suppressed yukawa couplings are more elusive to existing searches. In this section, we summarize the constraints on the charged and neutral Higgs bosons of Type I 2HDMs, with a focus on the light mass region.

\subsection{Indirect Bounds}
\label{sec:constraintsIndirect}


As previously mentioned, indirect bounds on light Type I 2HDMs are mild. Besides $b\rightarrow s\gamma$ already discussed in Sec.~\ref{sec:intro}, other competing constraints from flavor observables are $B_s\rightarrow\mu^+\mu^-$ and $\Delta M_{B_{s,d}}$, which are only marginally stronger, requiring that $\tanb\gsim 1.8-2.2$ in the mass range $m_{H^\pm}<m_t$ \cite{Enomoto:2015wbn}.

Another source of indirect constraints comes from contributions to the  electroweak oblique parameters, particularly $T$, induced by the mass splittings between $H^\pm$, $A^0$ and $H^0$. For the light spectra considered here, however, we have checked that $\Delta T$ constraints are easily evaded.

\subsection{Collider bounds}


LEP placed a robust lower bound on $m_{H^\pm}\gsim 78.6$~GeV by searching for the decay modes $H^+ \rightarrow c \bar{s},\, \tau^{+} \nu$, assuming the absence of any non-fermionic decays \cite{Searches:2001ac}. The DELPHI and OPAL collaborations also considered the bosonic decay $H^{\pm} \rightarrow W^{\pm*} A^{0}$~\cite{Abdallah:2003wd,Abbiendi:2008aa}. In Type I 2HDM scenarios, the $\tanb$-independent limits obtained were $m_{H^\pm}\gsim 77.6$ GeV (DELPHI) and $m_{H^\pm}\gsim 65$ GeV (OPAL), provided that $m_{A^{0}}\gsim12$~GeV.

The OPAL collaboration has searched for the associated production $e^+e^-\rightarrow A^0H^0$, with $A^0,\,H^0\rightarrow q\bar q,\,gg$, and $\tau^+\tau^-$ \cite{Abbiendi:2004gn,Abbiendi:2004ww}. While the resulting mass limits vary across the parameter space, they essentially turn off for either $A^0$ or $H^0$ heavier than $\sim 80\GeV$.

\begin{figure}[t]
\includegraphics[width=0.6\textwidth]{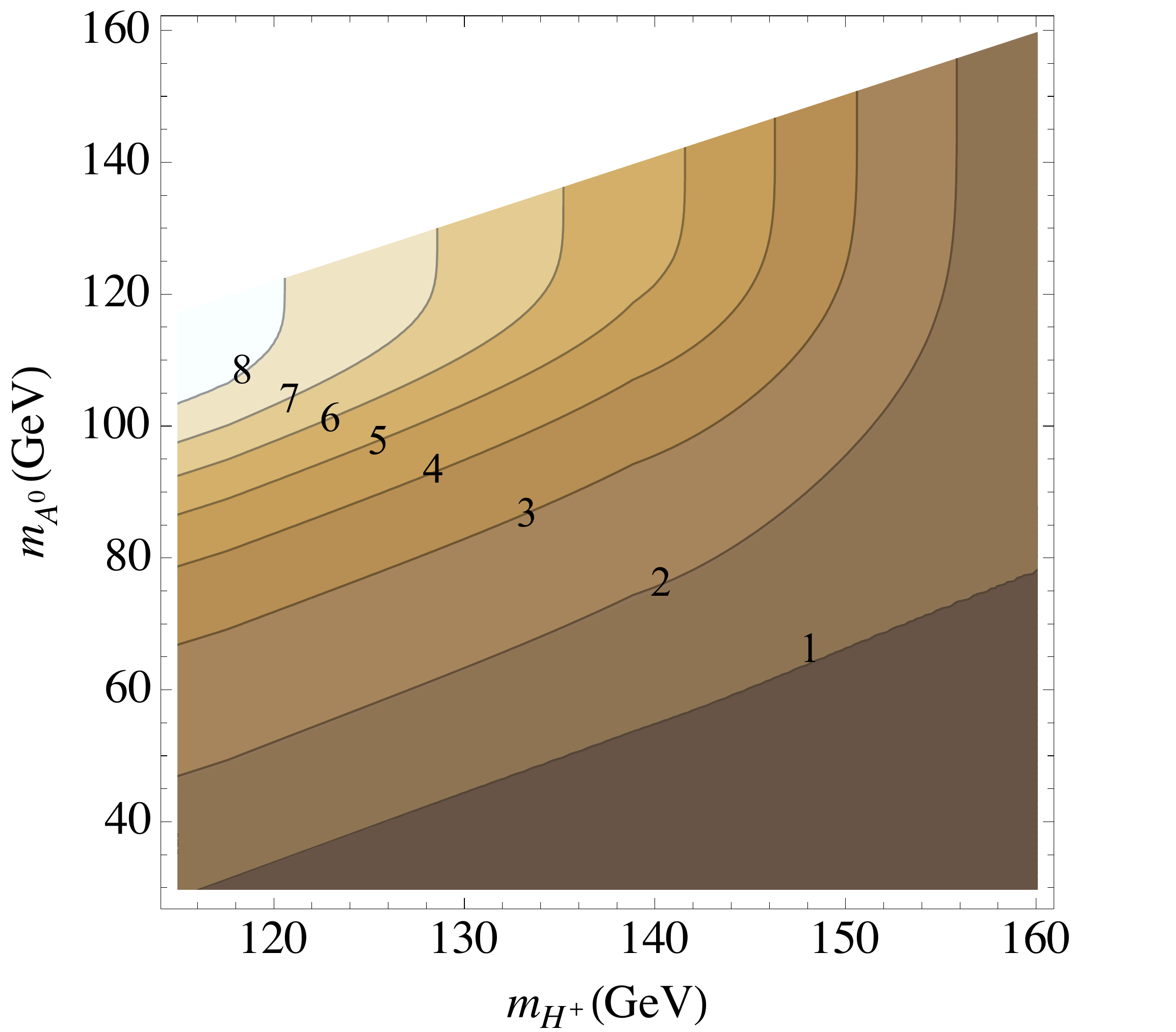}
	
	\caption{\label{fig:taunulimits} Reinterpretation of the 8 TeV CMS limits on $\text{Br}(t\rightarrow b\,H^+)\times\text{Br}(H^+\rightarrow \tau^+\nu)$ \cite{CMS:2014cdp} as lower bounds on $\tanb$ in a Type I 2HDM, assuming that $m_{H^0}>m_{H^\pm}$.}
\end{figure}

LEP, Tevatron and LHC searches for the SM Higgs are also potentially sensitive to the neutral scalars, $A^0$ and $H^0$. If $A^0,\,H^0$ are lighter than $\sim 110\GeV$, however, these states  decay dominantly to $b\bar b$ final states (with $\tau^+\tau^-$ as the subleading mode), and are challenging to probe at the Tevatron and LHC due to the large backgrounds and suppressed cross sections relative to the SM Higgs. While SM Higgs searches at LEP cannot constrain $A^0$ due to the absent $A^0Z^0Z^0$ coupling (see Eqs.~(\ref{eq:scalarcouplings1},\ref{eq:scalarcouplings2})), they are sensitive to $e^+e^-\rightarrow H^0Z^0$ production, and constrain $\zeta_{H}^2\lsim0.01-0.3$ in the range $m_{H^0}\simeq (15-115)\GeV$ \cite{Barate:2003sz}. 

Furthermore, LHC measurements of the SM Higgs properties provide non-trivial constraints on the parameters of the scalar potential (\ref{eq:higgspot}). If $\phi^0$ ($=A^0$ or $H^0$) is ligher than $m_{H_\text{SM}}/2$, the decay channel $\hsm\rightarrow \phi^0\phi^0$ can easily dominate the SM Higgs width for generic values of the quartic couplings in (\ref{eq:higgspot}). In order to avoid conflict with observations, in the mass range $m_{\phi^0}\lesssim62$~GeV the tri-scalar coupling $\lambda_{\phi\phi\hsm}$ must be suppressed, $\lambda_{\phi\phi\hsm}\lesssim (2-6)$~GeV. While this condition is not generically satisfied, it is a parameter that can be adjusted independently of the physical masses and mixing angles $\delta$ and $\beta$. Since the charged Higgs phenomenology we will consider is not directly affected by $\lambda_{\phi\phi\hsm}$, we will include the region $m_{\phi^0}\lesssim62$~GeV in our study, with the implicit assumption that $\Gamma(\hsm\rightarrow \phi^0\phi^0)$ is not in conflict with observations.

Another parameter that is directly constrained by SM Higgs measurements is $\delta$ -- current data pushes the model towards the alignment limit where $\delta$ is small and the properties of the 125 GeV Higgs are ``SM-like''. For a more thorough discussion on that, see \cite{Alves:2012ez,Bernon:2015qea,Bernon:2015wef,Haber:2017udj}.



Previously mentioned upper bounds on $\text{Br}(t\rightarrow b\,(H^+\rightarrow \tau^+\nu))$ from ATLAS and CMS are sensitive enough to be relevant even if the decay mode $H^+\rightarrow \tau^+\nu$ is subdominant, as is the case of the Type I 2HDMs we are considering. We recast the 8 TeV CMS constraints on the branching ratio $\text{Br}(t\rightarrow b\,H^+)\times\text{Br}(H^+\rightarrow \tau^+\nu)$ \cite{CMS:2014cdp} as lower bounds on the value of \tanb, displayed in Fig.~\ref{fig:taunulimits} as a function of $m_{H^\pm}$ and $m_{A^0}$, assuming for simplicity (and without loss of generality) that $m_{H^0}>m_{H^\pm}$.


To summarize, the light mass region of Type I 2HDMs is still experimentally viable in a vast swath of parameter space. In what follows, we investigate how this region can be constrained by existing LHC searches for the Standard Model $t\,\bar t\,H_{_\text{SM}}$ process.

\section{Signals in \tth ~Searches}
\label{sec:colliders}

\begin{figure}[t]
\includegraphics[width=0.7\textwidth]{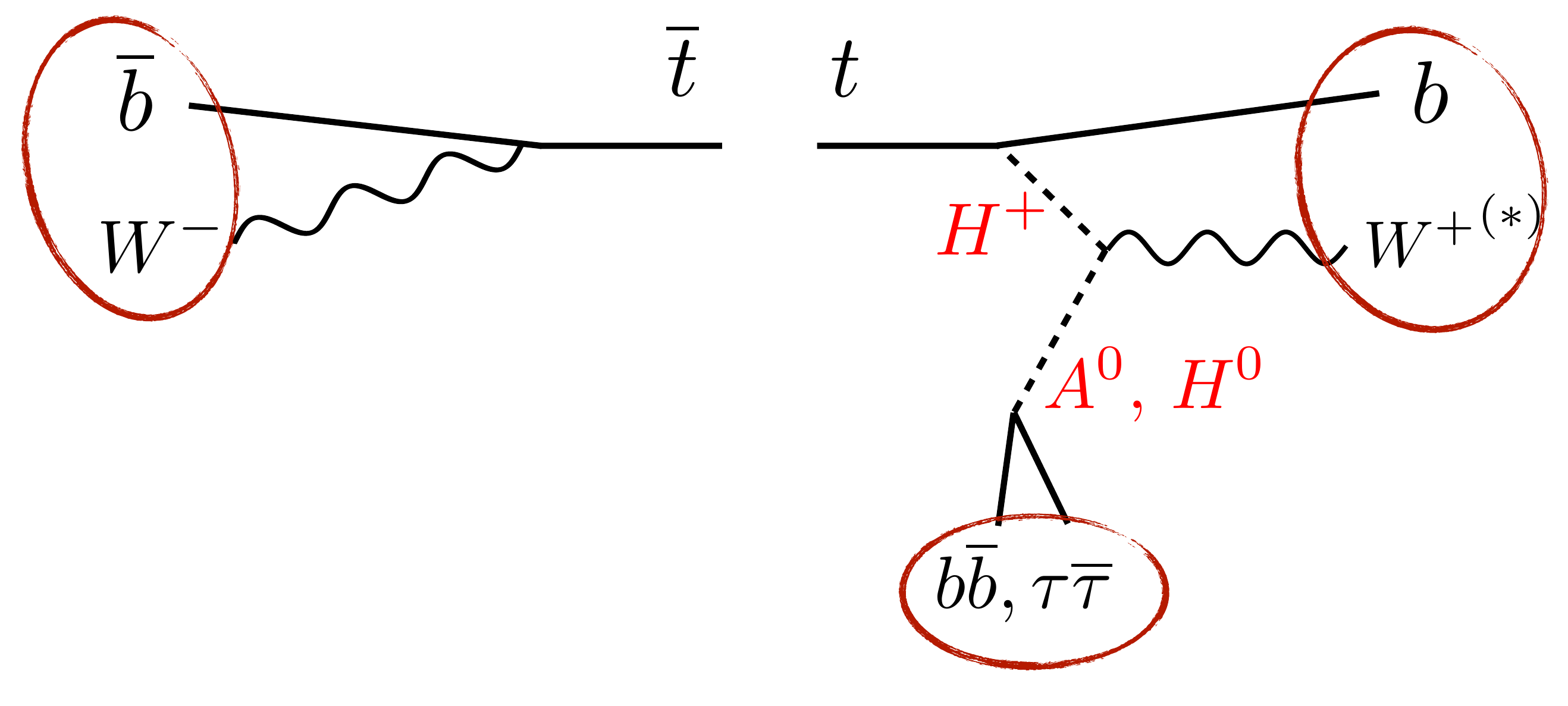}
	\caption{Signal from rare top decay to $b\,H^\pm$ in a light Type I 2HDM, whose final states overlap with those of SM \tth.}
\label{fig:topology}
\end{figure}

Although the parameter space of Type I 2HDMs with mass hierarchy:
\beq\label{massHierarchy}
M_{\phi^0=A^0\,\text{or}\,H^0}~<~~ M_{H^0_{_{\text{SM}}}}\;,\;M_{H^\pm}~~<~M_{t}
\eeq
is still experimentally viable, its phenomenology remains relatively unexplored in comparison to that of heavier 2HDM spectra. A stricking signal of models with (\ref{massHierarchy}) are rare top decays that can contaminate $\tth$ searches, particularly those targeting leptonic or $b\bar b$ decays of the SM Higgs, as illustrated in Fig.~\ref{fig:topology}. Given the very large top pair cross section, and the fact that $\text{Br}(t\rightarrow b H^+)$ can be as high as a few percent, this contamination
can lead to observable excesses in $\tth$ measurements relative to the SM expectation. The excess pattern, however, would appear inconsistent across different channels if interpreted as an enhanced $\tth$ signal strength. Generically, no excess should appear on $\gamma\gamma$ channels, since searches typically require $m_{\gamma\gamma}\approx 125$~GeV (within resolution) to specifically target the SM Higgs. On the other hand, one would expect excesses in channels targeting $b\bar b$ and $\tau^{+} \tau^{-}$ final states, albeit with different strengths. Many of these analyses use Multivariate discriminants (MVAs), such as boosted decisions trees, neural networks, etc., which may be tuned to the specific final state kinematics of $\tth$. In those cases, the contamination from rare top decays may be partially filtered out, depending on how well the MVAs can discriminate between the two processes. Normally, details of MVA based studies are not public, and the extraction of limits on contaminant signals is unfeasible.

In fact, SM Higgs searches as early as the Tevatron's could have been contaminated by rare top decays. The CDF collaboration has searched for $t\,\bar t\,(h^0\rightarrow b\,\bar b) $ over the mass range $m_{h^0}=(100-150)$~GeV~\cite{Collaboration:2012bk, Aaltonen:2013ipa}, and observed an $\mathcal{O}(2\sigma)$ excess above expectations at $m_{h^0}\sim(100-105)$~GeV. The reported best fit for the signal strength was $\mu_{t\bar t h^0}=7.40^{+4.65}_{-3.80}$\, for $m_{h^0}=100~\GeV$, and  $\mu_{t\bar t h^0}=8.56^{+4.82}_{-4.10}$\, for $m_{h^0}=105~\GeV$. This would correspond to a rate of roughly $(65\pm 40)$~fb, or, in terms of the inclusive top pair cross section, $(0.003 - 0.015)\times\sigma_{t\,\bar t}$.  This analysis was MVA based, and its mass resolution was limited due to the presence of four $b$-quarks in the signal final state, leading to a combinatoric ambiguity in identifying the $b$-jets originating from $h^0$ decays, and therefore to a broadening of the expected $m_{b\bar b}$ peak. All these factors preclude us from inferring any concrete implications regarding a potential contamination from BSM processes, but the results are nonetheless intriguing, and, if corroborated with further deviations at the LHC, could warrant a re-analysis of the Tevatron's data.

At the LHC, existing $\tth$ results from ATLAS and CMS do seem to suggest a pattern of excesses inconsistent with the hypothesis of enhanced $\tth$ production, although with small statistical significance. At 8 TeV, the uncertainties in $\tth$ measurements are too large to offer any indication of an excess or lack thereof \footnote{One notable exception is the CMS measurement in the same-sign dilepton channel \cite{Khachatryan:2014qaa}, which gives the following best fit for the signal strength: $\mu_{ttH}=5.3^{+2.1}_{-1.8}$\,.}, with a combined best fit of $\mu_{ttH}=2.3^{+0.7}_{-0.6}$ (ATLAS + CMS, all channels) \cite{Khachatryan:2016vau}. Existing 13 TeV data is inconclusive as well, showing no excess in CMS $b\bar b$ (MVA) \cite{CMS-PAS-HIG-16-038} and ATLAS $\gamma\gamma$ \cite{ATLAS:2016nke}, and $\mathcal{O}(1\sigma)$ excesses in CMS multileptons (MVA) \cite{CMS-PAS-HIG-16-022, CMS:2017vru},  CMS $\gamma\gamma$ \cite{CMS:2016ixj}, ATLAS $b\bar b$ (MVA) \cite{ATLAS-CONF-2016-080}, and ATLAS multileptons \cite{ATLAS-CONF-2016-058}. Of all 13 TeV searches to date, only the latter, ATLAS multileptons, employs a traditional cut-and-count procedure\footnote{The $4\ell$ category in the CMS multilepton analysis \cite{CMS:2017vru} employs a cut-and-count procedure as well, but the final measurement has uncertainties substantially larger than the ones in ATLAS, and therefore, is less sensitive.}, and therefore is amenable to recasting in terms of our charged Higgs signal. We shall do so in the following section.



We end by commenting on our choice of spectrum when reinterpreting the ATLAS multilepton results. As previously discussed, if (\ref{massHierarchy}) is realized, the charged Higgs will dominantly decay to $W^{(*)}\phi^0$, where $\phi^0$ is the {\it lightest} neutral scalar. Since LHC measurements of the 125~GeV Higgs push this model into the alignment limit, $(\sin\delta)^2\lsim 0.1$, the signals in Fig.~\ref{fig:topology} will be essentially independent of whether $\phi^0=A^0$ or $H^0$, since
\be
\frac{\text{Br}(H^\pm\rightarrow W^{\pm(*)}H^0)}{\text{Br}(H^\pm\rightarrow W^{\pm(*)}A^0)}\,\Bigg|_{m_{A^0}=m_{H^0}}=~~~1-(\sin\delta)^2.
\ee
Moreover, if lighter than $\sim 110$ GeV, $A^0$ and $H^0$ will have the same leading branching ratios, namely, $\text{Br}(\phi^0\rightarrow b\overline b)\approx 0.8$ and $\text{Br}(\phi^0\rightarrow \tau^+\tau^-)\approx 0.08$. For practical purposes, therefore, we will choose $A^0$ as the lightest neutral scalar and decouple $H^0$ (i.e., set $m_{H^0}>m_{H^\pm}$) for the remainder of the paper. None of the results that follow change in any significant way if $A^0 \rightarrow H^0$. The only loss of generality that comes with this assumption is the possibility of two independent decay modes. However, \cite{ATLAS-CONF-2016-058} does not rely on a mass peak, the efficiencies do not vary dramatically over the range of $m_{A^0,\,H^0}$ considered, and, in the presence of a mass hierarchy between $A^0$ and $H^0$, the lighter mode dominates the bosonic decays of $H^\pm$. Consequently, even this complication should not impact our results significantly.

\section{Recast of the ATLAS 13 TeV Search in Multileptons}
\label{sec:discussion}

\newcommand{\Am}{\cite{ATLAS-CONF-2016-058} }
\newcommand{\tauh}{\tau_\text{had} }

At the time of writing of this paper, \cite{ATLAS-CONF-2016-058} was the most recent ATLAS search for $\tth$ production in the multilepton channel, corresponding to $13.2~\text{fb}^{-1}$ of  13 TeV data. 
It targeted leptonic decays of the SM Higgs in $WW^*$, $\tau^+\tau^-$, and $ZZ^*$, by looking into four exclusive signal regions, namely, same-sign dileptons and one hadronically decaying $\tau$ ($2\ell1\tauh$), same-sign dileptons vetoing hadronically decaying $\tau$'s ($2\ell0\tauh$), 3 leptons ($3\ell$), and 4 leptons ($4\ell$). Rare top decays in Fig.~\ref{fig:topology}, with $A^0\rightarrow \tau^+\tau^-$, can contaminate all of these signal regions. We performed a Monte Carlo (MC) study to obtain a quantitative estimate of this contamination and the sensitivity of \Am to the Type I 2HDM spectra of (\ref{massHierarchy}). Overall, we were able to validate our MC simulation of \Am by reproducing the $\tth$ efficiencies provided by ATLAS to within a factor of 2 (12 independent efficiencies in total). This translates into a factor of $\sim2$ uncertainty in our results for the signal strength, and a $\sim50\%$ uncertainty in our results for $\tanb$ (since our signal scales as $\tanb^{-2}$). For a detailed description of our MC study, as well as our statistical treatment of fits and exclusions, we refer the reader to Appxs.~\ref{sec:recast}, \ref{appx:stats}. Throughout our study, we include the contribution of the SM $\tth$ process, assuming $\mu_{t\bar t H}=1$.

\begin{figure}[t]
\centering
\begin{subfigure}
\centering
\includegraphics[width=0.495\textwidth]{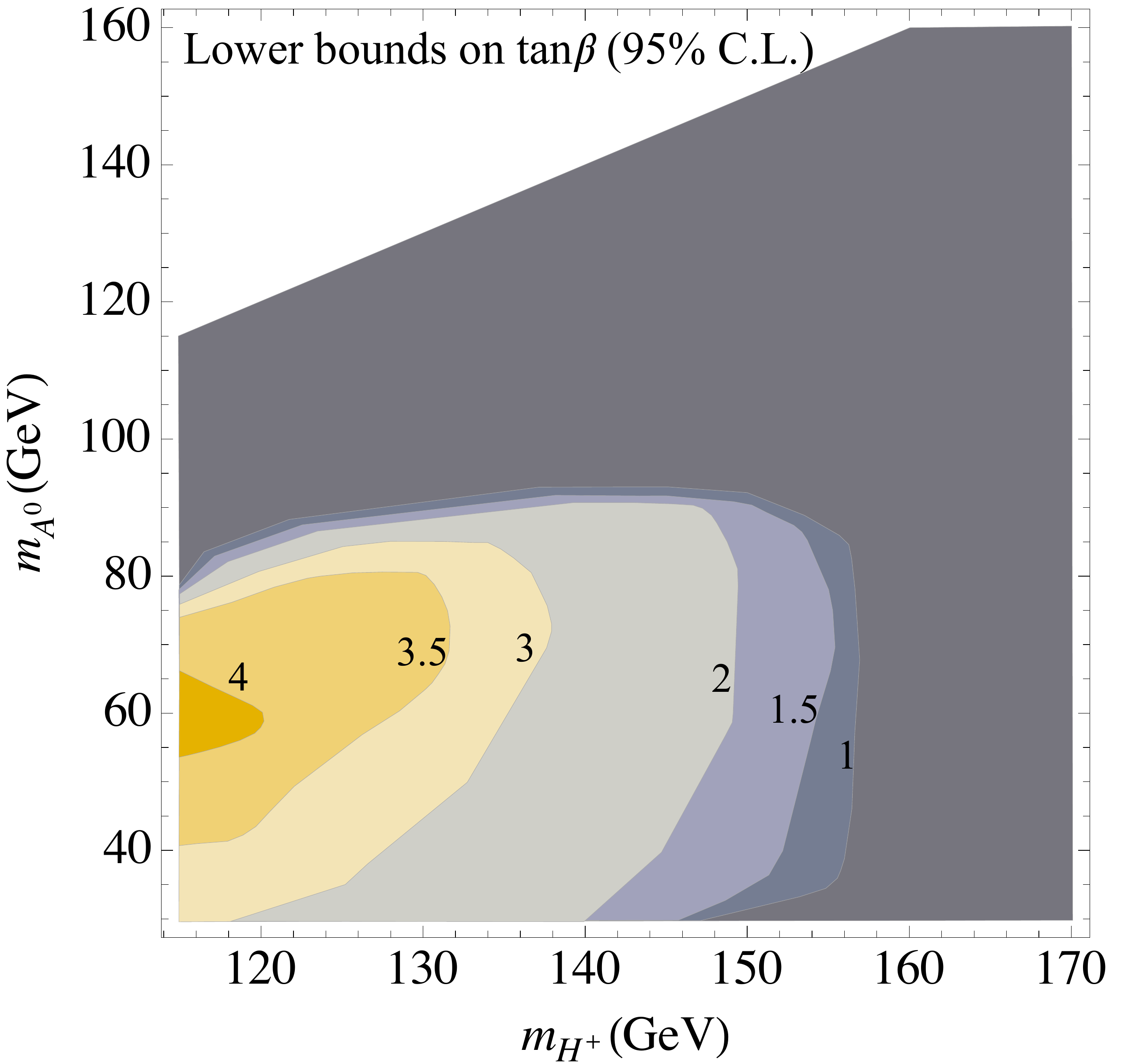}
\end{subfigure}%
\begin{subfigure}
\centering
\includegraphics[width=0.49\textwidth]{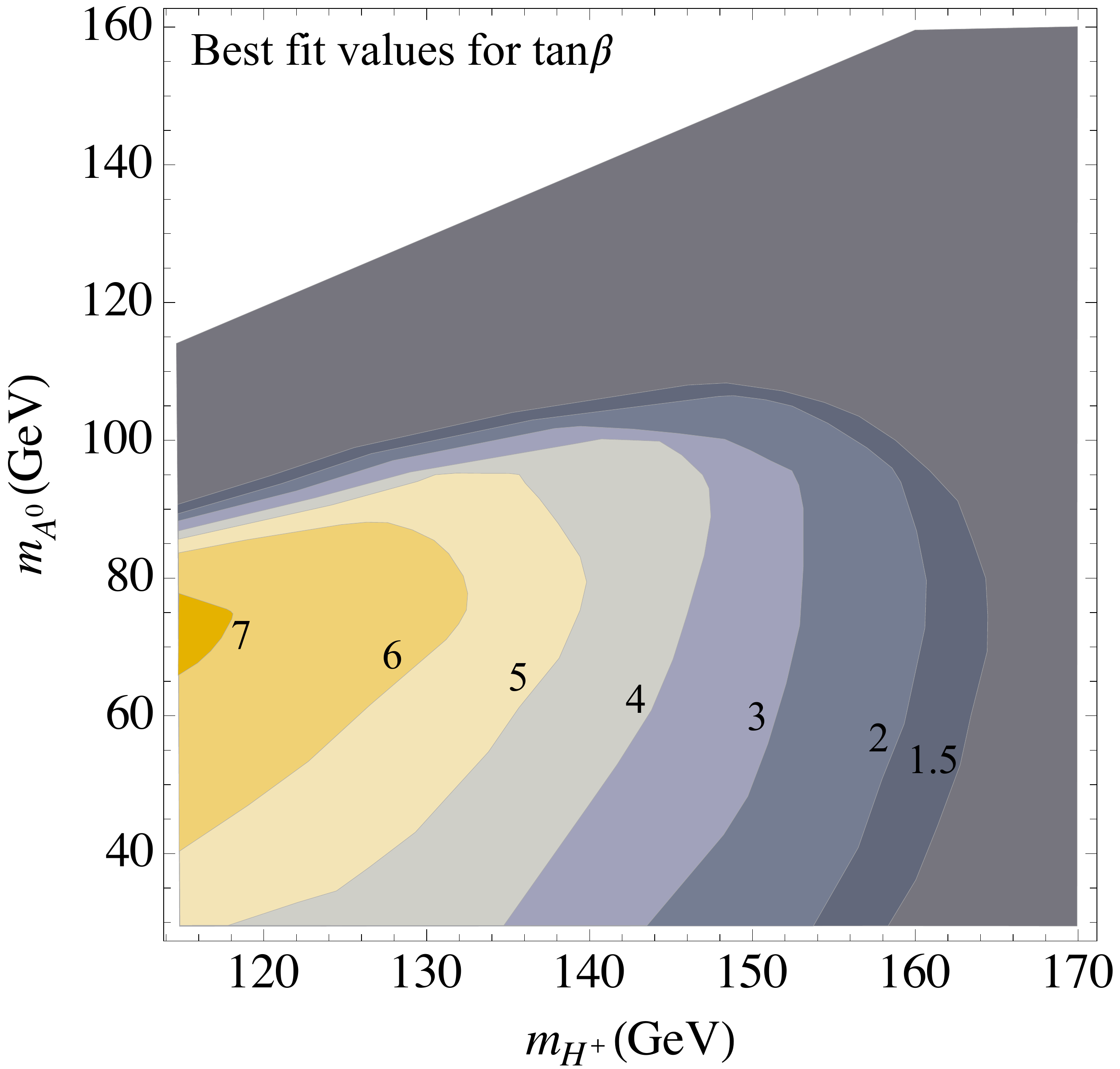}
\end{subfigure}
	\caption{Recast of the ATLAS $\tth$ search in multileptons. (Left) Lower bounds on $\tanb$ inferred from limits on $t \rightarrow b\,\hplus\rightarrow b\,W^{+*} (A^0\rightarrow \tau^+\tau^-)$. (Right) Values of $\tanb$ that best fit the data.}
\label{fig:tanbLimFit}
\end{figure}

We first obtain upper bounds on the contamination from $t\rightarrow b \hplus$, so that the expected number of events in the signal regions of \Am are compatible with observations. We interpret these constraints as a lower bound on $\tanb$ in the $m_\hplus ,\,m_\ahp$ mass plane, shown in Fig.~\ref{fig:tanbLimFit}(a). Our signal yield is suppressed in the compressed regions $m_{H^\pm} - m_\ahp\lsim30$~GeV (where $\text{Br}(H^\pm\rightarrow W^{\pm(*)}A^0)$ is small), and $m_t-m_{H^\pm}\lsim20$~GeV (where $\text{Br}(t\rightarrow b\,H^{+})$ is small). In these regions of suppressed signal yield, the inferred lower bounds on $\tanb$ are innocuous.


These limits would be stronger but for the presence of excesses in the data. Consequently, it behooves us to see whether we can understand these excesses as arising from rare top decays. In Fig.~\ref{fig:tanbLimFit}(b) we show the best fit for $\tanb$ at each point in the parameter space of $m_{H^+}$ and $m_{A^0}$. Notably, the efficiencies for the process $t \rightarrow b\,\hplus\rightarrow b W^+ (A^0\rightarrow \tau^+\tau^-)$ (before folding in branching ratios) do not vary dramatically across the $m_\hplus ,\,m_\ahp$ mass plane.

\begin{figure}[t]
\includegraphics[width=0.6\textwidth]{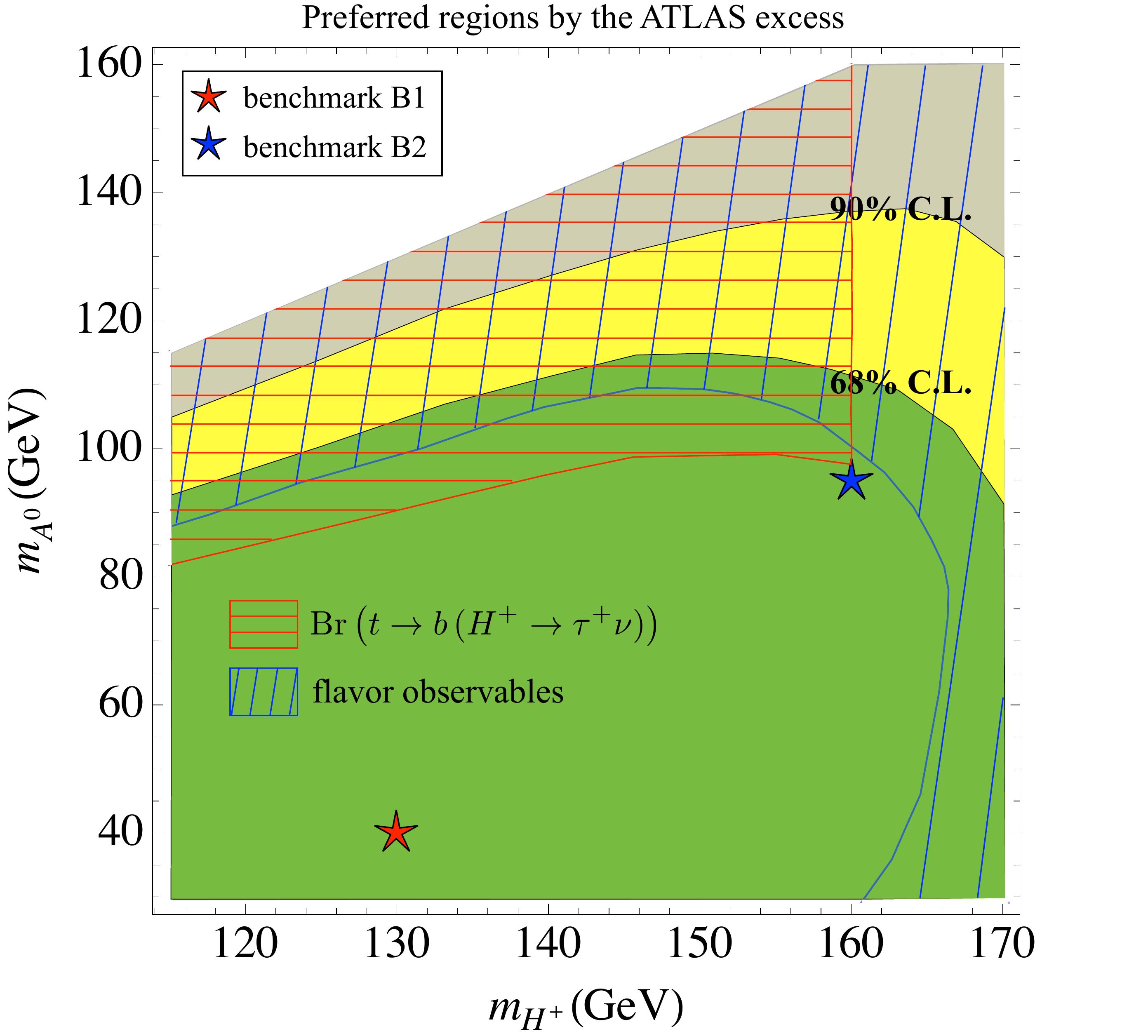}
	\caption{Preferred regions from the fit to the data in \cite{ATLAS-CONF-2016-058}, at 68\% C.L. (green region), and 90\% C.L. (yellow region). Also shown: regions in tension with indirect flavor observables, and in tension with CMS limits on $\text{Br}(t \rightarrow b\, (\hplus\rightarrow\tau^+\nu))$. These tensions only apply if the best $\tanb$ fit is assumed for each mass point.}
\label{fig:parameterspace}
\end{figure}

In Fig.~\ref{fig:parameterspace}, we show the preferred regions of parameter space, defined from the goodness of fit to the data in \Am. Here, the values of $\tanb$ are profiled at each mass point to yield the best fit (see Fig.~\ref{fig:tanbLimFit}(b)). Under this assumption, some regions of parameter space are excluded by other measuments. In particular, the compressed region $m_{H^\pm} - m_\ahp\lsim40$~GeV is in tension with the CMS bounds on $\text{Br}(t \rightarrow b\, (\hplus\rightarrow\tau^+\nu))$. Likewise, the compressed region $m_t-m_{H^\pm}\lsim10$~GeV is in tension with $b$ flavor observables discussed in Sec.~\ref{sec:constraintsIndirect}.

\begin{figure}[t]
\includegraphics[width=0.8\textwidth]{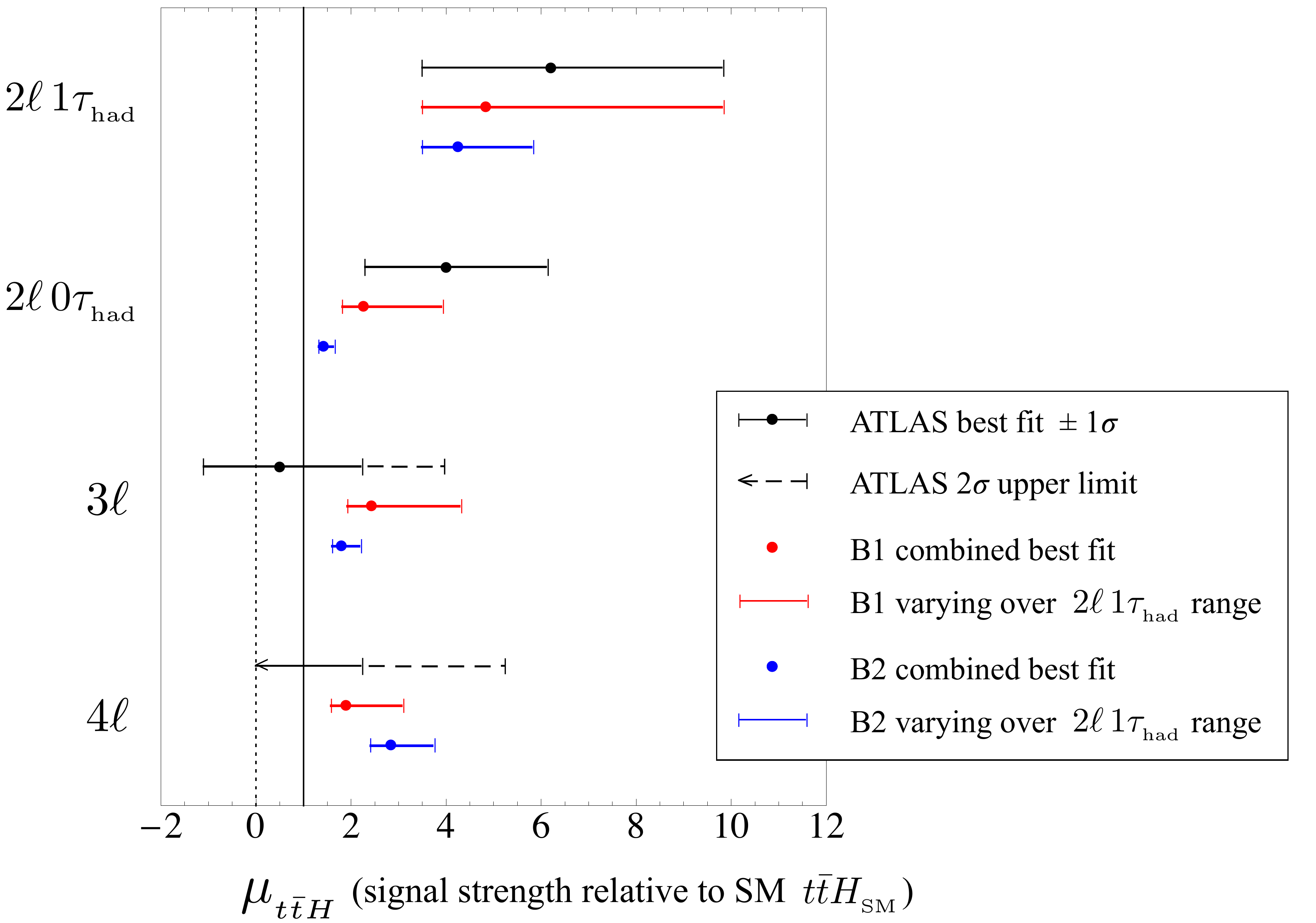}
	\caption{Signal strenght measured by ATLAS in each of the four signal regions, as well as the predictions from our two benchmark points, B1 and B2. The prediction from the best $\tanb$ fits of B1 and B2 are shown as red and blue dots, respectively.  Moreover, $\tanb$ is varied to encompass the $1\sigma$ range favored by data in the $2\ell1\tauh$ channel. Only values of $\tanb$ consistent with flavor constraints are used, leading to a shorter allowed range for B2. Note that the signal strengths displayed here are subject to up to a factor of 2 uncertainty, stemming from our MC estimation of signal efficiencies.}
\label{fig:mu}
\end{figure}

Finally, we select two benchmark points to illustrate the pattern of contamination across the four signal regions of \cite{ATLAS-CONF-2016-058}. The first benchmark point, B1, is at low mass, $m_{H^\pm}=130$~GeV, $m_\ahp=40$~GeV. For this benchmark, $H^\pm$ can decay to an on-shell $W^\pm$, yielding a harder charged lepton. Direct production of $H^+H^-$ and $H^\pm A^0$ is several hundreds of femtobarns at 13 TeV, and in principle dedicated searches in final states with 1 or 2 leptons plus 3 or more $b$-jets could be sensitive to this point.

The second benchmark, B2, is at high mass, and close to the excluded compressed regions, $m_{H^\pm}=160$~GeV, $m_\ahp=95$~GeV. Because of the small mass splitting between $t$ and $H^+$, the $b$-jet from $t\rightarrow b\,H^+$ is relatively soft and has a lower probability of passing the $b$-tagging requirements. Overall, B2 predicts that roughly 72\% of its signal in the $2\ell1\tauh$ region has only one $b$-tagged jet. Indeed, the data in the $2\ell1\tauh$ channel seem to indicate that the excess appears exclusively in the 1 $b$-tag category. Since current statistics is very low, this might be due to Poisson fluctuations, and only more data will be able to settle this matter. 

A final comment on B2 regards our choice of $m_{A^0}$ for this benchmark. This was motivated by an excess at $m_{H^0}\approx (90-100)$~GeV in the combination of LEP seaches for the SM Higgs  \cite{Barate:2003sz}, which could be interpreted in our scenario as a signal of $H^0$, if this were the lightest neutral scalar with $\zeta_H^2\sim\mathcal{O}(10^{-1})$ (see Eqs.(\ref{eq:scalarcouplings1}),(\ref{eq:scalarcouplings2})).

In Fig.~\ref{fig:mu}, we show the predicted signal strength from benchmarks B1 and B2, measured in units of the SM $\tth$ signal strength, for each of the four signal regions of \cite{ATLAS-CONF-2016-058}. Both the combined best fit, as well as a selected range of $\tanb$ are shown. Specifically, this range is chosen so it encompasses the $1\sigma$ range favored by data in the $2\ell1\tauh$ channel. We point out that for B2, it is not possible to reach the $1\sigma$ upper range of $2\ell1\tauh$ without violating flavor observables. For this reason, we cut off the range of B2 at this exclusion boundary.

We can see that the typical excess pattern seen by \Am is well explained by the hypothesis of contamination from rare top decays. At this point, however, the uncertainties are still large enough that even the no signal hypothesis is marginally consistent with observations. Upcoming analyses with more data will either tighten the exclusions of our model, or, optimistically, corroborate the deviations from the SM expectation.

\section{Discussion}
\label{sec:conclusion}

While the LHC has so far found no compelling evidence for new physics, tremendous possibilities still exist for discovery of new particles, including light electroweakly charged ones. 

In Type I two Higgs doublet models, a charged Higgs lighter than the top quark can naturally be produced at significant rates from rare top decays. Remarkably, if there are additional light neutral scalars in the spectrum, the final state for such signals has a substantial overlap with those of SM $\tth$ processes. As a consequence, it is natural to consider signals from light extended Higgs sectors as a contaminant to existing SM searches.

Interestingly, many - but not all - of the existing \tth\ searches show excesses, both at the Tevatron as well as the 8 and 13 TeV runs of the LHC. It is challenging to simultaneously reconcile these excesses with each other - significant excesses in leptonic channels, and an inconclusive pattern in $\gamma \gamma$ and $b\bar b$ channels, for instance. If these excesses persist, they could potentially be explained by the contamination of a new, charged Higgs signal from top decays.

On general grounds, one would expect that the more tuned a given analysis is to the specific final state kinematics of the $\tth$ process, the less sensitive it should be to BSM top decays. That would be the case of LHC searches for $t\bar t(\hsm\rightarrow \gamma \gamma)$, as these focus on a narrow $m_{\gamma \gamma}$ window around 125 GeV; or of (post Higgs discovery) MVA analyses in general. On the other hand, more inclusive analyses should be more prone to contamination from signals of extended Higgs sectors. Examples of more inclusive analyses are the Tevatron's CDF search for $t\bar t (h^0\rightarrow b\bar b)$ \cite{Collaboration:2012bk, Aaltonen:2013ipa}, which considers a broader $m_{b\bar b}$ window of $100 - 150$~GeV; and the 13 TeV ATLAS search for leptonic $\tth$ \cite{ATLAS-CONF-2016-058}, which employs a more conventional cut-and-count strategy, and might have non-negligible acceptance to BSM signals in multileptons plus two or more $b$-jets.

We have recast the 13 TeV ATLAS search for $\tth$ in multilepton final states \cite{ATLAS-CONF-2016-058}, and found that its signal regions can be naturally contaminated by a light Type I 2HDM spectrum at low $\tanb$.
Our recast of the results of \cite{ATLAS-CONF-2016-058} provides new limits on these models. Furthermore, these models can also explain the excesses observed in the data without exceeding null results from other measurements. 
In principle, this signal could also show up in other searches, such as those targeting $\hsm\rightarrow b\bar b$, depending on the details of the MVA used, and the masses of  $\ahp/H^0$. Indeed, considering the high branching ratio of $\ahp/H^0 \rightarrow b \bar b$, final states with many $b$-jets may provide the strongest tests going forward.

Should the excesses persist, it is clear that light Type I 2HDM spectra provide an attractive potential explanation. Broadening  search windows, especially for $\ahp/H^0 \rightarrow b \bar b$ at lower $m_{b \bar b}$, could further constrain this scenario, or, possibly, provide the first evidence at the LHC of physics beyond the Standard Model.

\begin{acknowledgments}
We thank A. Djouadi and M. Spira for answering our questions regarding charged Higgs 3-body decays, and Kyle Cranmer for helpful discussions on statistics. We also thank A. Pierce and M. Graesser for interesting conversations about the ATLAS $ttH$ excess. NW and DA are supported by the NSF under grants PHY-0947827 and PHY-1316753. SEH is supported by the Cluster of Excellence Precision Physics, Fundamental Interactions and Structure of Matter (PRISMA-EXC 1098) and by the Mainz Institute for Theoretical Physics.

\end{acknowledgments}

\begin{appendix}
\section{Recasting the ATLAS search for $\tth$ in multileptons}
\label{sec:recast}
\subsection{Overview}
The production of a Higgs boson in association with two top quarks is expected in the SM and has therefore been searched for by ATLAS and CMS~\cite{ATLAS-CONF-2016-080,CMS-PAS-HIG-16-038,ATLAS-CONF-2016-058,CMS-PAS-HIG-16-022}. In our study, we focus on the ATLAS $13$~TeV search in multilepton final states \cite{ATLAS-CONF-2016-058}, since it is based on a traditional cut-and-count analysis that is possible to recast. This search is targeted at events where the Higgs decays to either $WW^*$, $ZZ^*$, or $\tau^+\tau^-$. All these channels can lead to signatures with up to four leptons, for which the backgrounds are extremely low at the LHC. The potentially interesting events are grouped into four different signal regions: two with a pair of same-sign light leptons and either zero or one hadronic tau ($2\ell 0\tau$ and $2\ell 1\tau$), one with three light leptons ($3\ell$) and one with four light leptons ($4\ell$). The $13.2~\text{fb}^{-1}$ data have shown an excess in both the $2\ell 1\tau$ and the $2\ell 0\tau$ regions, more particularly for events with only one $b$-tagged jet. The peculiar structure of this excess cannot be explained solely by an enhanced $\tth$ cross section and could be a sign for new physics.

Naively, the $(bW)(bW^*\tau\overline{\tau})$ final state associated with our Type I 2HDMs is similar to the $t\overline{t}(\hsm\rightarrow \tau\overline{\tau})$ signal that is looked for in this search. The kinematics of the final states is however significantly different. Notably, only one $W$ boson arises from the direct decay of a top quark. The other one is produced through the decay of the charged Higgs to $\hhp/\ahp$ and is therefore much softer or even off-shell. Consequently, when the mass splitting between $H^+$ and $\hhp/\ahp$ is small, events with $4$ leptons will have lower efficiencies for trigger and preselection cuts. Similarly, the efficiency of our signal in the $3\ell$ region should be lower than that of $t\overline{t}\hsm$. The observed rates for the $2\ell 0\tau$ and $2\ell 1\tau$ regions should however remain significant. Broadly speaking, this structure is similar to the excess observed in ATLAS.

\subsection{The recasting procedure}

In the ATLAS $\tth$ search, events are selected and classified into four exclusive signal regions associated with various cuts on the multiplicity and momenta of leptons, hadronic taus, light jets and $b$-jets. In addition to these cuts, vetoes against $m_{\ell\ell}\simeq m_{Z}$ as well as low mass leptonic Drell-Yan are imposed in events with same-flavor lepton pairs. Depending on the signal region, the light flavor leptons can also be required to verify specific isolation criteria. Although most of the selection cuts are straightforward, the tagging of hadronic taus as well as the isolation requirements on the light leptons require a more careful treatment that we detail in what follows.  

\begin{table}[t]
    \begin{tabular}{|c|| c|c|c|c|}
	\hline
	& $~2\ell 0\tau~$ & $~2\ell 1 \tau~$ & $~~3\ell~~$ & $~~4\ell~~$ \\
	\botrule
	$N_{\text{PGS}}$ & 1.2 & 1.4 & 1.7 & 0.20 \\
	\hline
	$N_{\text{Delphes}}$ & 1.0 & 1.1 & 2.1 & 0.55 \\
	\hline
	$N_\text{ATLAS}$ & 1.7 &  0.73 & 1.2 & 0.11\\
	\hline
    \end{tabular}
    \caption{\label{tab:validation1} Expected signal yield from $t\bar t (\hsm\rightarrow \tau\tau)$ in the signal regions of \cite{ATLAS-CONF-2016-058}, obtained using \texttt{Delphes} and \texttt{PGS} for simulation of the detector response, as well as the expected yield provided by ATLAS.}
\end{table}

In the $2\ell 0\tau$, $2\ell 1\tau$ and $3\ell$ regions, the light leptons have to pass loose and tight isolation cuts. Since these cuts are associated with efficiencies of $99$\% and $96$\% respectively~\cite{Aad:2016jkr}, we do not implement them in our analysis. In the $4\ell$ region, the electrons and muons are submitted to so-called ``gradient'' isolation cuts, with a $p_T$-dependent efficiency. These cuts can have a significant impact at low $p_T$ and we therefore take them into account when using \texttt{Delphes}~\cite{deFavereau:2013fsa} for detector simulation.

To estimate the errors associated with our modeling of the detector response, we use two different detector simulators: \texttt{Delphes 3} and \texttt{PGS~4}~\cite{deFavereau:2013fsa,Conway:2008pgs}. In \texttt{Delphes}, we set the $b$- and $\tau$-tagging efficiencies/mistag rates to the values used by ATLAS in~\cite{ATLAS-CONF-2016-058} and initially loosen the electron and muon isolation criteria. The rest of the parameters are set to the default values given in the ATLAS $13$~TeV card from \texttt{Delphes}. In order to take into account the possible loss of low $p_T$ electrons or muons in the $4\ell$ signal region, we select the light leptons in this region with efficiencies corresponding to the ones of the gradient isolation cuts.  When generating events with \texttt{PGS}, we use the default lepton tagging algorithm and do not apply any additional isolation cuts. For hadronic taus, however, since the efficiency of the \texttt{PGS} tagging algorithm is much lower than the one used by ATLAS, we modify the \texttt{PGS} code to identify all jets within $\Delta R = 0.2$ of a parton-level tau as hadronic taus.
We apply a flat tau-tagging efficiency  {\it ex post facto}, given by the branching-ratio-weighted average of the one-prong and three-prong efficiencies given in \cite{ATLAS-CONF-2016-058}.
Likewise, we modify the \texttt{PGS} $b$-tagging algorithm to better represent the working point used in \cite{ATLAS-CONF-2016-058}.
All the other cuts besides the preselection cuts are implemented without any change.

In order to validate our analysis, we generate $\tth$ events using \texttt{MadGraph5}~\cite{Alwall:2011uj}, and study each of the Higgs decay modes ($\tau^+\tau^-$,  $WW^*$ and $ZZ^*$) independently. We match these events up to one additional jet and shower them with \texttt{Pythia6}~\cite{Sjostrand:2006za} using MLM matching~\cite{Hoche:2006ph} with the $k_T$ shower scheme~\cite{Alwall:2014hca}. As mentioned above, we use both \texttt{PGS 4}~\cite{Conway:2008pgs} and \texttt{Delphes 3}~\cite{deFavereau:2013fsa} to simulate the detector response. The expected event yields obtained by our MC study for each of the three Higgs decay channels are given in tables~\ref{tab:validation1}, \ref{tab:validation2} and \ref{tab:validation3}. We observe that \texttt{Delphes} and \texttt{PGS} exhibit complementary performances. \texttt{Delphes} gives a better modeling of the ATLAS efficiencies in the $2\ell 1\tau$ region, whereas \texttt{PGS} shows better agreement in the $2\ell 0\tau$, $3\ell$ and $4\ell$ regions.
Overall, \texttt{PGS} provides a better modeling of the SM $\tth$ efficiencies, and therefore we use \texttt{PGS} for recasting the results in \cite{ATLAS-CONF-2016-058} in terms of rare top decays to $b\,H^{+}$.
In our analysis, we use the recasting procedure as is. That is, we do not apply any {\it ad hoc} correction factors to the efficiencies, since we could not determine exactly the origin of our small discrepancies. 
The overall uncertainties in our detector simulation translate into a factor of $\sim 2$ uncertainty in our results for the signal yields.

\begin{table}[t]
    \begin{tabular}{|c||c|c|c|c|}
	\hline
	& $~2\ell 0\tau~$ & $~2\ell 1 \tau~$ & $~~3\ell~~$ & $~~4\ell~~$ \\
	\botrule
	$N_{\text{PGS}}$ & 5.8 & 1.1 & 4.8 & 0.42 \\
	\hline
	$N_{\text{Delphes}}$ & 6.7 & 0.80 & 8.4 & 1.3 \\
	\hline
	$N_\text{ATLAS}$ & 7.5 &  0.66 &4.6  & 0.42\\
	\hline
    \end{tabular}
    \caption{\label{tab:validation2} Expected signal yield from $t\bar t (\hsm\rightarrow WW^*)$ in the signal regions of \cite{ATLAS-CONF-2016-058}, obtained using \texttt{Delphes} and \texttt{PGS} for simulation of the detector response, as well as the expected yield provided by ATLAS.}
\end{table}

\begin{table}
    \begin{tabular}{|c||c|c|c|c|}
	\hline
	& $~2\ell 0\tau~$ & $~2\ell 1 \tau~$ & $~~3\ell~~$ & $~~4\ell~~$ \\
	\botrule
	$N_{\text{PGS}}$ & 0.25 & 0.06 & 0.17 & 0.021 \\
	\hline
	$N_{\text{Delphes}}$ & 0.37 & 0.069 & 0.59 & 0.16 \\
	\hline
	$N_\text{ATLAS}$ & 0.29 &  0.03 & 0.25  & 0.05\\
	\hline
    \end{tabular}
    \caption{\label{tab:validation3} Expected signal yield from $t\bar t (\hsm\rightarrow ZZ^*)$ in the signal regions of \cite{ATLAS-CONF-2016-058}, obtained using \texttt{Delphes} and \texttt{PGS} for simulation of the detector response, as well as the expected yield provided by ATLAS.}
\end{table}

\begin{figure}[t]
    \includegraphics[width=0.3\linewidth]{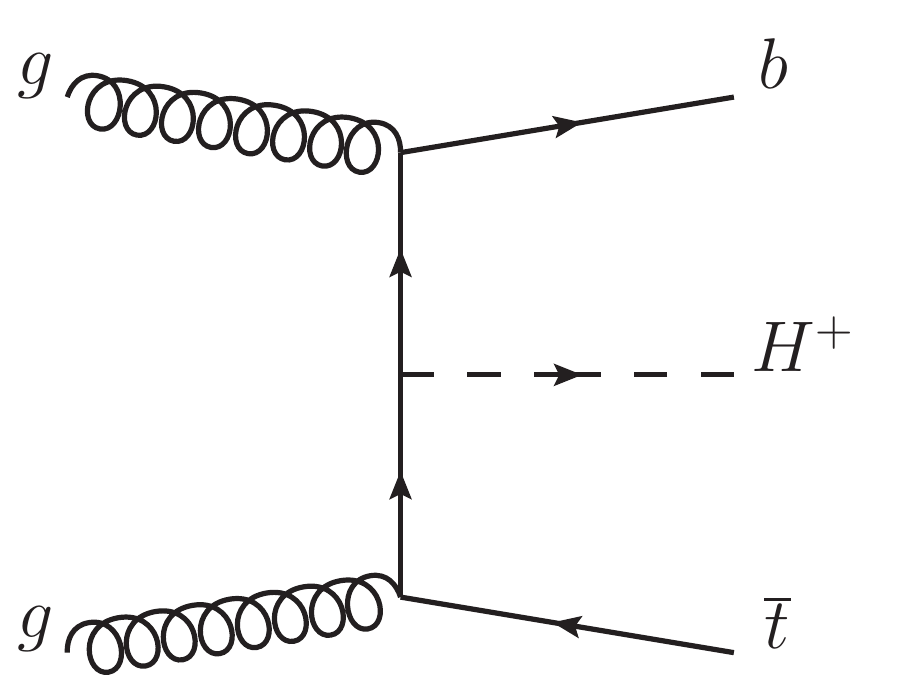}
    \caption{\label{fig:offshell} Feynman diagram for non-resonant production of $H^+\bar{t}b$.}
\end{figure}

\subsection{Signal generation}
A charged Higgs can be produced along with a top quark and a bottom quark through two distinct processes. First, as highlighted throughout this paper, this final state can arise from on-shell top pair-production, with one of the tops decaying to $H^+ b$. As the charged Higgs gets heavier than about $155$~GeV, however, the branching ratio $\text{Br}(t\rightarrow b\,H^+)$ gets suppressed, and off-shell production of the charged Higgs must be included as it gives an important contribution to the signal rate. The dominant non-resonant $H^\pm$ production process, $pp\rightarrow \bar t\,b\,H^+$, is shown in Fig.~\ref{fig:offshell}. We include this process in our study for all charged Higgs masses above $150$~GeV, and use the NLO cross sections provided in \cite{Degrande:2016hyf}.

We use \texttt{MadGraph}~\cite{Alwall:2011uj} to generate $t\bar{t}+j$ events, and to decay both tops to $(\bar b W^-)(b\,H^+)$ (and the corresponding charge conjugate process). We match these events up to one additional jet and shower them using \texttt{Pythia}~\cite{Sjostrand:2006za} with MLM matching~\cite{Hoche:2006ph} and the $k_T$ shower scheme~\cite{Alwall:2014hca}. We use \texttt{Pythia} to further decay the $W$'s, $H^\pm$'s and $A^0$'s. For the detector simulation, we use 
\texttt{PGS}~\cite{Conway:2008pgs} with the same settings as described above. The non-resonant process  $tbH^\pm\rightarrow(\bar b W^-)bH^\pm$ and its charge conjugate are generated with \texttt{MadGraph}. The subsequent steps are the same as for the first process.

\section{Signal yields and statistical procedure}
\label{appx:stats}

\subsection{Signal yields}

As discussed in Appx.~\ref{sec:recast}, we use MC to obtain the signal efficiencies in each of the four signal regions defined by ATLAS, for both the $t\bar t$ on-shell process:
\be
pp\rightarrow\bar t\,t \rightarrow \bar t\,b\, (H^+\rightarrow W^{+(*)}\, ( A^0\rightarrow \tau^+\tau^-))\,,
\ee
 as well as the non-resonant process in Fig.~\ref{fig:offshell}
 \be
pp\rightarrow \bar t\,b\,  (H^+\rightarrow W^{+(*)}\, ( A^0\rightarrow \tau^+\tau^-))\,.
 \ee
We generically denote the respective efficiencies by $\epsilon_\text{res}$ and $\epsilon_\text{nonres}$.

The signal yield for a specific signal region is then given by:
\be
S= S_\text{res} + S_\text{nonres}\,,
\ee
where
\be
S_\text{res} = \epsilon_\text{res}\times\left(\sigma_{t\bar t}\times\mathcal{L}\right)\times\left(2\,\text{Br}(t\rightarrow b \, H^+)\times\text{Br}(H^+\rightarrow W^{+(*)}\,A^0)\times\text{Br}(A^0\rightarrow \tau^+\tau^-)\right)\,,
\ee
and
\be
S_\text{nonres} = \epsilon_\text{nonres}\times\left(\frac{\sigma_{\text{nonres}}}{\tanb^2}\times\mathcal{L}\right)\times\left(\text{Br}(H^+\rightarrow W^{+(*)}\,A^0)\times\text{Br}(A^0\rightarrow \tau^+\tau^-)\right)\,.
\ee

For the total cross sections, we use $\sigma_{t\bar t}\,\big|_{\text{13 TeV}}=830$~pb, and the following approximate fit for $\sigma_{\text{nonres}}$ extracted from \cite{Degrande:2016hyf}:
\be
\sigma_{\text{nonres}}(m_{H^\pm})\,\Big|_{\text{13 TeV}} = \left(22.53 - 0.106\;\frac{m_{H^\pm}}{\text{GeV}}   \right)\;\text{pb}
\ee
in the range $m_{H^\pm}\sim[150-175]$~GeV.

\subsection{Branching ratios}

Throughout our study we set $\text{Br}(A^0\rightarrow \tau^+\tau^-)=0.083$.

The top branching ratio to $b\,H^+$ is given by \cite{Gunion:1989we}:
\be
\text{Br}(t\rightarrow b \, H^+)\,=\,\frac{ R_{H^+}}{\;1\,+\, R_{H^+}}\,,
\ee
where
\beq
 R_{H^+}~&&=~\frac{\Gamma(t\rightarrow b \, H^+)}{\Gamma(t\rightarrow b \, W^+)} \nonumber\\
&&=~ \frac{p_{H^+}}{p_{W^+}}\,\frac{1}{\tanb^2}\,\frac{(m_t^2+m_b^2-m_{H^+}^2)(m_t^2+m_b^2)-4\,m_b^2\,m_t^2}{m_W^2\,(m_t^2+m_b^2-2\,m_W^2)+(m_t^2-m_b^2)^2}\,.
\eeq
Above, $p_{H^+}$ is the momentum of $H^+$ in the top's rest frame,
\be
p_{H^+}\;=~\frac{m_t}{2}\,\sqrt{\left(1-\frac{m_{H^+}^2}{m_t^2}\right)^2  -2\;\frac{m_b^2}{m_t^2}\,\left(1+\frac{m_{H^+}^2}{m_t^2}\right)+\;\frac{m_b^4}{m_t^4}\, }\;,
\ee
and $p_{W^+}$ is defined in an analogous way. 

The branching ratios of the charged Higgs are computed from its various widths. For completeness, we list all of them below.

The charged Higgs width to a pair of {\it light} on-shell SM fermions is given by:
\be
\Gamma(H^+\rightarrow f \bar f^\prime) ~=~ \frac{G_F}{4\sqrt{2}\pi}\,\frac{m_{H^+}}{\tanb^2}\;N_c\,|U_{f\bar f^\prime}|^2\,\left( m_f^2\,+\,m^2_{\bar f^\prime} \right)\,,
\ee
where $N_c=3$ and $U_{f\bar f^\prime}$ is an element of the CKM matrix if $f,\bar f^\prime$ are quarks, and $N_c=1$ and $U_{f\bar f^\prime}=1$ otherwise;
$m_f$, $m_{\bar f^\prime}$ above are the {\it running masses} at $\mu=m_{H^+}$.

Since we are interested in spectra where $m_{H^+}<m_t$, the width $\Gamma(H^+\rightarrow t^*\, \bar b)$ goes via an off-shell top quark, and is given by \cite{Djouadi:1995gv}:
\beq
\Gamma(H^+\rightarrow W^+b\,\bar b)~=~&&\frac{3\,G_F^2\,m_t^4}{64\,\pi^3}\,\frac{m_{H^+}}{\tanb^2}\,\bigg(\frac{\kappa_W^2}{\kappa_t^3}\,(4\kappa_W\kappa_t+3\kappa_t-4\kappa_W)\,\text{log}\left(\frac{\kappa_W\,(\kappa_t-1)}{\kappa_t-\kappa_W}\right)  \nonumber\\
&&+ ~(3\kappa_t^2-4\kappa_t-3\kappa_W^2+1)\,\text{log}\left(\frac{\kappa_t-1}{\kappa_t-\kappa_W}\right) -\frac{5}{2} \\
&& +~\frac{1-\kappa_W}{\kappa_t^2}\,(3\kappa_t^3-\kappa_t\kappa_W-2\kappa_t\kappa_W^2+4\kappa_W^2) + \kappa_W\,(4-3\kappa_W/2 )   \bigg)\, , \nonumber
\eeq
where $\kappa_t=m_t^2/m_{H^+}^2$, and $\kappa_W=m_W^2/m_{H^+}^2$. Above, we neglect the contribution from the bottom yukawa coupling.

The bosonic widths of the charged Higgs, $\Gamma(H^+\rightarrow W^{+(*)}\,\phi^0)$, depend on whether the final state $W^+$ is on- or off-shell. In the former case, we have:
\be
\Gamma(H^+\rightarrow W^{+}\,\phi^0)~=~(1-\zeta_\phi^2)\,\frac{G_F}{8\sqrt{2}\,\pi}\,\frac{m_W^4}{m_{H^+}}\,\sqrt{\lambda(m_{\phi^0}^2,m_W^2,m_{H^+}^2)}\,\lambda(m_{\phi^0}^2,m_{H^+}^2,m_W^2)\, ,
\ee
where $\zeta_\phi$ for $\phi^0 = \hsm,\, H^0,\,A^0$ are given in Sec.~\ref{sec:model}, and
\be
\lambda(x,y,z)~=~\left(1-\frac{x}{z}-\frac{y}{z}  \right)^2-4\,\frac{x\,y}{z^2}.
\ee
On the other hand, if the $W^{+}$ is off-shell, we have \cite{Djouadi:1995gv}:
\be
\Gamma(H^+\rightarrow W^{+*}\,\phi^0)~=~(1-\zeta_\phi^2)\;m_{H^+}\,\frac{9\, G_F^2\,m_W^4}{8\,\pi^3}\,G\left(\frac{m_{\phi^0}^2}{m_{H^+}^2},\frac{m_W^2}{m_{H^+}^2}\right)\,,
\ee
where
\beq
G(x,y)~&&=~\frac{1}{8}\,\Bigg( 2\,(-1-x+y)\,\sqrt{\lambda_G(x,y)}\,\bigg(\frac{\pi}{2}+\text{ArcTan}\bigg[\frac{y\,(1+x-y)-\lambda_G(x,y)}{(1-x)\sqrt{\lambda_G(x,y)}}  \bigg]   \bigg)  \nonumber\\
&&+~\big(\lambda_G(x,y)-2x\big)\,\text{log}(x) + \frac{(1-x)}{3\,y}\,\big(5\,y\,(1+x)-4\,y^2+2\,\lambda_G(x,y) \big)\Bigg)\,,
\eeq
and
\be
\lambda_G(x,y)~=\, -1+2\,x+2\,y-(x-y)^2\,.
\ee

\subsection{Statistical treatment of fits and exclusions}

For each signal region in \cite{ATLAS-CONF-2016-058} with $N_\text{sr}$ observed events, $\mu_{B\,\text{sr}} \pm\Delta B_\text{sr}$ expected background events, and $S_\text{sr}$ expected signal events, we define the likelihood function:
\be
\mathcal{L}_{\text{sr}}(S_\text{sr}, B_\text{sr})=P(N_\text{sr}\,|\,S_\text{sr}+B_\text{sr})\times G(B_\text{sr}\,|\,\mu_{B\,\text{sr}}\,,\,\Delta B_\text{sr})\,,
\ee
where $P$ and $G$ are Poisson and Gaussian distributions, respectively. The likelihood for the combination of all four signal regions is given by the product of the individual likelihoods:
\be
\mathcal{L}(\mu_S, \theta_B)=\prod_{\text{srj}}\mathcal{L}_{\text{srj}}(S_\text{srj},B_\text{srj})\,,
\ee
where $\theta_B\, =\, (B_\text{sr1},\,B_\text{sr2},\,B_\text{sr3},\,B_\text{sr4})$, and $\mu_S=(m_{H^+}, m_{A^0}, \tanb)$ uniquely specifies a point in the model parameter space, and unambiguously determines the signal yield $S_\text{srj}$ in each signal region. Since the correlations between background uncertainties in the four signal regions were not provided in \cite{ATLAS-CONF-2016-058}, we treat these uncertainties as uncorrelated. We note, however, that the post-fit backgrounds in \cite{ATLAS-CONF-2016-058} did not change substantially relative to their pre-fit counterparts. This, combined with the factor of 2 uncertainties in our MC modeling of signal efficiencies, leads us to expect that the our results would not change substantially were we to properly include all background correlations in our fits. 

For obtaining limits, we use a profiled log-likelihood analysis. First, we define:
\be
\lambda(\mu_S)~=~\text{log}\,\mathcal{L}(\mu_S, \hat{\hat{\theta}}_B) - \text{log}\,\mathcal{L}(\hat{\mu}_S, {\hat{\theta}}_B)\,,
\ee
where the unconditional likelihood estimators $\hat{\mu}_S, \,{\hat{\theta}}_B$ maximize the global $\text{log}\,\mathcal{L}(\mu_S, \theta_B)$, and the conditional estimator $\hat{\hat{\theta}}_B$ maximizes $\text{log}\,\mathcal{L}(\mu_{S}, \theta_B)$ for a given $\mu_{S}$. Since there are $3$ independent degrees of freedom in ${\mu}_S$, namely, $m_{H^+}$, $m_{A^0}$, and $\tanb$, the $p$-value for a specific model defined by ${\mu}_S$ is determined by:
\be
p~=~1-\text{CDF}(\chi^2_3,\,-2\,\lambda(\mu_S)),
\ee
where $\text{CDF}(\chi^2_3,\,-2\,\lambda(\mu_S))$ is the cumulative distribution function for a $\chi^2$-distribution with 3 degrees of freedom, evaluated at $-2\,\lambda(\mu_S)$. From the $p$-value we can determine the exclusion confidence level for all points in the studied parameter space.

Finally, we note that when finding the goodness of fit of a given mass point $\mu^\prime_S=(m_{H^+}, m_{A^0})$, we profile over the value of $\tanb$ that maximizes the log-likelihood, i.e., we use
\be
\lambda(\mu^\prime_S)~=~\text{log}\,\mathcal{L}(\mu^\prime_S,\,  \text{tan}\hat{\hat{\beta}},\, \hat{\hat{\theta}}_B) - \text{log}\,\mathcal{L}(\hat{\mu}^\prime_S,\,  \text{tan}{\hat{\beta}},\, {\hat{\theta}}_B)\,.
\ee
In this case, we define the $p$-value in an analogous way,
\be
p~=~1-\text{CDF}(\chi^2_2,\,-2\,\lambda(\mu^\prime_S)),
\ee
but instead use a $\chi^2$-distribution with only 2 degrees of freedom.

\end{appendix}

\bibliography{REFS}

\end{document}